# Quantifying synergistic information using intermediate stochastic variables

Rick Quax[1,a], Omri Har-Shemesh[1], Peter M.A. Sloot[1,2,3]

1. Computational Science Lab, University of Amsterdam, The Netherlands;
2. Advanced Computing Lab, ITMO University, Saint Petersburg, Russia;
3. Complexity Institute, Nanyang Technological University, Singapore.

a. Part of this work was presented at the Conference of Complex Systems 2016.

**Quantifying synergy among stochastic variables is an important open problem in information theory. Information synergy occurs when multiple sources together predict an outcome variable better than the sum of single-source predictions. It is an essential phenomenon in biology such as in neuronal networks and cellular regulatory processes, where different information flows integrate to produce a single response, but also in social cooperation processes as well as in statistical inference tasks in machine learning. Here we propose a metric of synergistic entropy and synergistic information from first principles. The proposed measure relies on so-called synergistic random variables (SRVs) which are constructed to have zero mutual information about individual source variables but non-zero mutual information about the complete set of source variables. We prove several basic and desired properties of our measure, including bounds and additivity properties. In addition, we prove several important consequences of our measure, including the fact that different types of synergistic information may co-exist between the same sets of variables. A numerical implementation is provided, which we use to demonstrate that synergy is associated with resilience to noise. Our measure may be a marked step forward in the study of multivariate information theory and its numerous applications.**

## 1  Introduction

Shannon's information theory is a natural framework for studying the correlations among stochastic variables. Claude Shannon proved that the entropy of a single stochastic variable uniquely quantifies how much information is required to identify a sample value from the variable, which follows from four quite plausible axioms (non-negativity, continuity, monotonicity and additivity) [1]. Using similar arguments, the mutual information between two stochastic variables is the only pairwise correlation measure which quantifies how much information is shared. However, higher-order informational measures among three or more stochastic variables remain a long-standing research topic [2–6].

A prominent higher-order informational measure is synergistic information [3–5,7–10], however it is still an open question how to measure it. It should quantify the idea that a set of variables taken together can convey more information than the summed information of its individual variables. Synergy is studied for instance in the context of regulatory processes in cells and networks of neurons. To illustrate the idea at a high level, consider the recognition of a simple object, say a red square, implemented by a multi-layer neuronal network. Some input neurons will implement local edge detection, and some other input neurons will implement local color detection, but the presence of the red square is not defined solely by the presence of edges or red color alone: it is defined as a particular higher-order relation between edges and color. Therefore, a neuronal network which successfully recognizes an object must integrate the multiple pieces of information in a synergistic manner. However, it is unknown exactly how and where this is implemented in any dynamical network because no measure exists to quantify synergistic information among an arbitrary number of variables.

We consider the task of predicting the values of an outcome variable $Y$ using a set of source variables $X \equiv \{X_i\}_i$. The total predictability of $Y$ given $X$ is quantified information-theoretically by the classic Shannon mutual information,

$$I(X:Y) = H(Y) - H(Y \mid X).$$

Here,

$$H(Y) = -\sum_y \Pr(Y=y) \log_2 \Pr(Y=y)$$

is the entropy of $Y$ and denotes the total amount of information needed to determine a unique value of $Y$, in bits. It is also referred to as the uncertainty about $Y$. The conditional variant $H(Y \mid X)$ obeys the chain rule $H(X,Y) = H(X) + H(Y \mid X)$ and is written explicitly as

$$H(Y \mid X) = -\sum_x \Pr(X=x) \sum_y \Pr(Y=y \mid X=x) \log_2 \Pr(Y=y \mid X=x).$$

This denotes the remaining entropy of $Y$ given that the value for $X$ is observed.

In this article we address the problem of quantifying synergistic information between $X$ and $Y$. To illustrate information synergy, consider the classic example of the XOR-gate of two i.i.d. binary inputs, defined by the following (deterministic) input-output table.

| $X_1$ | $X_2$ | $Y$ |
|---|---|---|
| 0 | 0 | 0 |
| 0 | 1 | 1 |
| 1 | 0 | 1 |
| 1 | 1 | 0 |

*Table 1: Transition table of the binary XOR-gate.*

*A priori* the outcome value of $Y$ is 50/50 distributed. It is easily verified that observing both inputs $X_1$ and $X_2$ simultaneously fully predicts the outcome value $Y$, while observing either input individually does not improve the prediction of $Y$ at all. Indeed, we find that:

$$I(X_1:Y) = 0,$$
$$I(X_2:Y) = 0,$$
$$I(X_1, X_2:Y) = 1.$$

In words this means that in this case the information about the outcome is not stored in either source variable individually, but is stored *synergistically* in the combination of the two inputs. In this case $Y$ stores whether $X_1 \neq X_2$, which is independent of the individual values of either $X_1$ or $X_2$.

Two general approaches to quantify synergy exist in the current literature. On the practical side, methods have been devised to approximate synergistic information using simplifying assumptions. An intuitive example is the 'whole minus sum' (WMS) method [10] which simply subtracts the sum of pairwise ('individual') mutual information quantities from the total mutual information, i.e., $I(X:Y) - \sum_i I(X_i:Y)$. This formula is based on the assumption that the $X_i$ are uncorrelated; in the presence of correlations this measure may become negative and ambiguous.

On the theoretical side, the search is ongoing for a set of necessary and sufficient conditions for a general synergy measure to satisfy. To our knowledge, the most prominent systematic approach is the Partial Information Decomposition framework (PID) proposed by Williams and Beer [3]. Here, synergistic information is implicitly defined by additionally defining so-called 'unique' and 'shared' information; together they are required to sum up to the total mutual information $I(X:Y)$, among other conditions. However, it appears that the original axioms of Shannon's information theory are insufficient to uniquely determine the functions in this decomposition framework [11], so two approaches exist: extending or changing the set of axioms [3,7,8,12], or finding 'good enough' approximations [3,6,9,10].

Our work differs crucially from the abovementioned approach. In fact, we will define 'synergy' from first principles which is incompatible with PID. We use a simple example to motivate our perceived incongruence of PID. Our proposed procedure of calculating synergy is based upon a newly introduced notion of perfect 'orthogonal decomposition' among stochastic variables. We will prove important basic properties which we feel any successful synergy measure should obey, such as non-negativity and insensitivity to reordering subvariables. We will also derive a number of intriguing properties, such as an upper bound on the amount of synergy that any other variable can have about a given set of variables. Finally, we provide a numerical implementation which we use for experimental validation and demonstrate that synergistic variables have increased resilience to noise, which is an important property at large and specifically in biological systems.

## 2 Results

### 2.1 Preliminaries

#### 2.1.1 Definition 1: orthogonal decomposition.

Following the intuition from linear algebra we refer to two stochastic variables $A, B$ as *orthogonal* in case they are independent, i.e., $I(A:B) = 0$. Given a joint distribution of two stochastic variables $A, B$ we say that a function $D(B, A) = B^\perp, B^\parallel$ is an *orthogonal decomposition* of $B$ with respect to $A$ in case it satisfies the following five properties:

$$\text{sufficiency: } I(B^\perp, B^\parallel : B) = H(B),$$
$$\text{orthogonality: } I(B^\perp : A) = 0,$$
$$\text{parallelism: } I(B^\parallel : A) = I(B : A), \quad (1)$$
$$\text{non-spuriousness: } I(B^\parallel : A | B) = 0,$$
$$\text{parsimony: } I(B^\parallel : B) = I(B : A).$$

In words, $B$ is decomposed into two orthogonal stochastic variables $B^\perp, B^\parallel$ so that (*i*) the two parts taken together are informationally equivalent to $B$; (*ii*) the orthogonal part has zero mutual information about $A$; and (*iii*) the parallel part has the same mutual information with $A$ as the original variable $B$ has.

### 2.1.2 Related literature on decomposing correlated variables

Our notion of orthogonal decomposition is related to the ongoing study of 'common random variable' definitions dating back to around 1970. In particular our definition of $B^\parallel$ appears equivalent to the definition by Wyner [13], here denoted $B_W^\parallel$, in the specific case that $I(B_W^\parallel : A, B) = I(B : A)$. That is, in Appendix 6.6 we show that under this condition their $B_W^\parallel$ satisfies all three requirements in Eq. (1) which do not involve $B^\perp$ (which remains undefined in Wyner's work). In brief, their $B_W^\parallel$ is the 'smallest' common random variable which makes $A$ and $B$ conditionally independent, i.e., $B_W^\parallel = \arg\min_{A \to W \to B} I(A, B : W)$. They show that $I(B_W^\parallel : A) \geq I(B : A)$. In case their minimization is not able to reach the condition $I(B_W^\parallel : A) = I(B : A)$ it is an open question whether this implies that our $B^\parallel$ does not exist for the particular $A, B$. The required minimization step to calculate $B_W^\parallel$ is highly non-trivial and solutions are known only for very specific cases [14,15].

To illustrate a different approach in this field, Gács and Körner [16] define their common random variable as the 'largest' random variable which can be extracted deterministically from both $A$ and $B$ individually, i.e., $f(A) = g(B) = B_0^\parallel$ for functions $f$ and $g$ chosen to maximize $H(B_0^\parallel)$. They show that $I(B_0^\parallel : A) \leq I(B : A)$ and it appears in practice that typically the 'less than' relation actually holds. Their variable is more restricted than ours but has applications in zero-error communication and cryptography.

### 2.1.3 Sufficiency of decomposition

Our definition of orthogonal decomposition is sufficient to be able to define a consistent measure of synergistic information, as we will show in Section 2.2. However we leave it as an open question whether Eq. (1) is actually more stringent than strictly necessary. Therefore, our statement is that if orthogonal decomposition is possible then our synergy measure is valid; in case it is not possible then it remains an open question whether this implies that it is impossible to define synergy in a similar manner to ours. Important future work is thus to try to minimize the conditions of orthogonal decomposition while leaving a synergy measure intact.

### 2.1.4 Satisfiability of decomposition

Indeed it turns out that it is not always possible to achieve a perfect orthogonal decomposition according to Eq. (1), depending on $A$ and $B$. For example, we demonstrate in Appendix 6.5 that for the case of binary-valued $A$ and $B$ it is impossible to achieve the decomposition in case $B$ depends on $A$ as $\Pr(B = A) = p_b$.

On the other hand, one sufficient condition for being able to achieve a perfect orthogonal decomposition is being able to restate $A$ and $B$ as $A = (W, X)$ and $B = (W, Y)$ for $W, X, Y$ independent from each other. In this case it is easy to see that $B^\parallel = W$ and $B^\perp = Y$ are a valid orthogonal decomposition. Such a restating could be reached by reordering and relabeling of variables and states.

We propose the following line of reasoning to (asymptotically) reach this restating of $A$ and $B$ or at least approximate it. Nevertheless the remainder of the paper simply assumes the existence of the orthogonal decomposition and does not use any particular method to achieve it.

Consider the Karhunen-Loève transform (KLT) [17–19] which can restate any stochastic variable $X$ as

$$X \stackrel{\text{KLT}}{=} \mu + \sum_{k=1}^{\infty} \alpha_k \cdot Z_k.$$

Here, $\mu$ is the mean of $X$, the $Z_k$ are pairwise independent random variables, and the coefficients $\alpha_k$ are real scalars. This transform could be seen as the random variables analogy to the well-known principle component analysis or the Fourier transform.

Typically this transform is defined for a range of random variables in the context of a continuous stochastic process $\{X^t\}_{a \leq t \leq b}$. Here each $X^t$ is decomposed by $Z_k$ which are defined through the $X^t$ themselves as

$$Z_k = \int_a^b \alpha_k(t) \cdot X^t dt.$$

Here, the coefficients scalars become functions $\alpha_k(t)$ on $[a,b]$ which must be pairwise orthogonal (zero inner product) and square-integrable. Otherwise the abovementioned transform applies to each single $X^t$ in the same way, now with $t$-dependent coefficients $\alpha_k(t)$. Nevertheless, for our purpose we leave it open how the $Z_k$ are chosen; through being part of a stochastic process or otherwise. We also note that the transform works similarly for the discrete case, which is often applied to image analysis.

Let us now choose a single sequence of $Z_k$ as our variable 'basis'. Now consider two random variables $A$ and $B$ which can both be decomposed into $Z_k$ as the sequences $\{\alpha_k Z_k\}$ and $\{\beta_k Z_k\}$, respectively. In particular, the mutual information $I(A:B)$ must be equal before and after this transform. Then the desired restating of $A$ and $B$ into $A = (W, X)$ and $B = (W, Y)$ is achieved by

$$\begin{aligned} W &= \{Z_k : \alpha_k > 0 \wedge \beta_k > 0\}, \\ X &= \{Z_k : \alpha_k > 0 \wedge \beta_k = 0\}, \\ Y &= \{Z_k : \alpha_k = 0 \wedge \beta_k > 0\}. \end{aligned} \quad (2)$$

The choice of the common $Z_k$ could either be natural, such as a common stochastic process of which both $A$ and $B$ are part, or a known common signal which two receivers intermittently record. Otherwise $Z_k$ could be found through a numerical procedure to attempt a numerical approximation, as is done for instance in image analysis tasks.

## 2.2 Proposed framework

### 2.2.1 Synergistic random variable

Firstly we define that $S$ is a *synergistic random variable* (SRV) of $X \equiv \{X_i\}_i$ if it satisfies the conditions:

$$I(S:X) > 0,$$
$$\forall i : I(S:X_i) = 0. \quad (3)$$

In words, an SRV stores information about $X$ as a whole but no information about any individual $X_i$ which constitute $X$. Each SRV $S_i$ is defined by a conditional probability distribution $\Pr_i(S_i|X)$ and is thus conditionally independent of any other SRV given $X$. We denote the collection of all possible, non-redundant SRVs of $X$ as the joint random variable $\sigma(X)$. We sometimes refer to $\sigma(X)$ as a set because the ordering of its marginal distributions (SRVs) is irrelevant due to their conditional independence.

### 2.2.2 Maximally synergistic random variables

The set $\sigma(X)$ may in general be uncountable, and many of its members may have extremely small mutual information with $X$, which would prevent any practical use. Therefore we introduce the notion of *maximally synergistic random variables* (MSRV) which we will also use in some proofs. We do not have a proof that this set is countable, however our numerical results (see especially Figure 3) show that a typical MSRV has substantial mutual information with $X$ (about 75% of the maximum possible). This suggests that either the set of MSRVs is countable or that the mutual information of a subset of MSRVs rapidly converges to their maximum, aiding a practical use.

We define the set of MSRVs of $X$, denoted $\Sigma(X)$, as the smallest possible subset of $\sigma(X)$ which still makes $\sigma(X)$ redundant, i.e.,

$$\Sigma(X) = \min_{|\Sigma|} \{\Sigma \subseteq \sigma(X) : H(\sigma(X)|\Sigma(X)) = 0\}. \quad (4)$$

Here, $|\Sigma|$ denotes the cardinality of set $\Sigma$ which is minimized. Intuitively, one could imagine building $\Sigma(X)$ by iteratively removing an SRV $S_i$ from $\sigma(X)$ in case it is completely redundant given another SRV $S_j$, i.e., if $\exists j : H(S_i|S_j) = 0$. The result is a set $\Sigma(X)$ with the same informational content (entropy) as $\sigma(X)$ since only redundant variables are discarded. In case multiple candidates for $\Sigma(X)$ would exist then any choice among them will induce the same synergy quantity in proposed our synergy measure.

### 2.2.3 Synergistic entropy of $X$

We interpret $\Sigma(X)$ as representing all synergistic information that any stochastic variable could possibly store about $X$. Therefore we define the *synergistic entropy* of $X$ as $H(\Sigma(X))$. This will be the upper bound on the synergistic information of any other variable about $X$.

### 2.2.4 Orthogonalized SRVs

In order to prevent doubly counting synergistic information we orthogonalize all MSRVs. Let us denote $(S_i)_i \equiv S_1, S_2, ...$ for a particular ordering of all MSRVs in $\Sigma(X)$. Then we convert $\Sigma(X)$ into a set of *orthogonal* MSRVs, or OSRVs for short, for a given ordering:

$$\Sigma^{\perp}_{(S_i)_i}(X) \equiv \left(S_i^{\perp} \in D(S_i; S_1, ..., S_{i-1})\right)_i. \tag{5}$$

In words, we iteratively take each MSRV $S_i$ in $\Sigma(X)$ and add its orthogonal part $S_i^{\perp}$ to the set $\Sigma^{\perp}_{(S_i)_i}(X)$ in the given order $(S_i)_i$. As a result, each OSRV in $\Sigma^{\perp}_{(S_i)_i}(X)$ is completely independent from the others in this set. $\Sigma^{\perp}_{(S_i)_i}(X)$ is still informationally equivalent to $\Sigma(X)$ because during its construction we only discard completely redundant variables $S_i^{\parallel}$ given other SRVs.

Note that each orthogonal part $S_i^{\perp}$ is an SRV if $S_i$ is an SRV (or MSRV), which follows from the contradiction of the negation: if $S_i^{\perp}$ is not an SRV then $\exists j: I(S_i^{\perp} : X_j) > 0$ and consequently $(S_i^{\parallel}, S_i^{\perp}) \notin \sigma(X)$ which contradicts $S_i \in \sigma(X)$ since $(S_i^{\parallel}, S_i^{\perp}) = S_i$ by the above definition of orthogonal decomposition.

### 2.2.5 Total synergistic information

We define the total amount of synergistic information that $Y$ stores about $X$ as:

$$I_{\text{syn}}(X \to Y) \equiv \max_{(S_i)_i} \sum_{S_i^{\perp} \in \Sigma^{\perp}_{(S_i)_i}(X)} I(Y : S_i^{\perp}). \tag{6}$$

In words, we propose to quantify synergy as the sum of the mutual information that $Y$ contains about each MSRV of $X$, after first making the MSRVs independent and then reordering them to maximize this quantity. In the next Section we will prove several desired properties which this definition satisfies; here we finish with an informal outline of the intuition behind this definition and refer to corresponding proofs where appropriate.

#### 2.2.5.1 *Outline of intuition of the proposed definition*

Our initial idea was to quantify synergistic information directly as $I(Y : \sigma(X))$, however we found that this results in undesired counting of non-synergistic information which we demonstrate in Section 2.4.3 and in Appendix 6.2.1. That is, two or more SRVs taken together do not necessarily form an SRV, meaning that their combination may store information about individual inputs. For this reason we use the summation over OSRVs. Intuitively, each term in the sum quantifies a 'unique' amount of synergistic information which none of the other terms quantifies, due to the independence among all OSRVs in $\sigma^{\perp}_{(S_i)_i}(X)$. That is, no synergistic information is doubly counted, which we also argue in Appendix 6.2 by proving that $I_{\text{syn}}(X \to Y)$ never exceeds $I(Y : \sigma^{\perp}_{(S_i)_i}(X))$. On the other hand, no possible type of synergistic information is ignored (undercounted). This can be seen from the fact that only fully redundant variables are ever discarded in the above process; also we prove for example in Section 2.3.6 in the sense that for any arbitrary $X$ there exists a $Y$ such that $I_{\text{syn}}(X \to Y)$ equals the maximum $H(\sigma(X))$, namely $Y = X$.

This summation is sensitive to the ordering of the orthogonalization of the SRVs. The reason for maximizing over these orderings is the possible presence of synergies among the SRVs themselves. We prove that $I_{syn}(X \to Y)$ handles correctly such 'synergy-among-synergies', i.e., does not lead to over counting or undercounting, in Appendix 6.3.

## 2.3 Basic properties

Here we first list important minimal requirements that the above definitions obey. The first four properties typically appear in the related literature either implicitly or explicitly as desired properties; the latter two properties are direct consequences of our first principle to use SRVs to encode synergistic information. The corresponding proofs are straightforward and sketched briefly.

### 2.3.1 Non-negativity

$$I_{syn}(X \to Y) \geq 0. \tag{7}$$

This follows from the non-negativity of the underlying mutual information function, making every term in the sum of Eq. (6) non-negative.

### 2.3.2 Upper-bounded by mutual information

$$I_{syn}(X \to Y) \leq I(X:Y). \tag{8}$$

This follows from the Data-Processing Inequality [20], where $X$ is first processed into $\Sigma^{\perp}_{(S_i)_i}(X)$ and then $I_{syn}(X \to Y) \leq I\left(\Sigma^{\perp}_{(S_i)_i}(X):Y\right)$ follows because we can write

$$\begin{aligned}
I\left(\Sigma^{\perp}_{(S_i)_i}(X):Y\right) &= H\left(S_1^{\perp},...,S_n^{\perp}\right) - H\left(S_1^{\perp},...,S_n^{\perp}|Y\right) \\
&= \sum_i H\left(S_i^{\perp}\right) - H\left(S_1^{\perp},...,S_n^{\perp}|Y\right) \\
&\geq \sum_i H\left(S_i^{\perp}\right) - \sum_i H\left(S_i^{\perp}|Y\right) \\
&= \sum_i I\left(S_1^{\perp}:Y\right) \\
&= I_{syn}(X \to Y).
\end{aligned} \tag{9}$$

Here, $S_i^{\perp}$ is understood to denote the $i^{th}$ element in $\Sigma^{\perp}_{(S_i)_i}(X)$ after maximizing the sequence $(S_i)_i$ used to construct $\Sigma^{\perp}_{(S_i)_i}(X)$ for computing $I_{syn}(X \to Y)$.

### 2.3.3 Equivalence class of reordering in arguments

$$I_{syn}\left((X_i)_i \to (Y_i)_j\right) = I_{syn}\left((X_{i'})_{i'} \to (Y_{i'})_{j'}\right), \tag{10}$$
for any reordered labelings $i'$ and $j'$.

This follows from the same property of the underlying mutual information function and that of the sum in Eq. (6).

### 2.3.4 Zero synergy about a single variable

$$I_{syn}(X_1 \to Y) = 0. \tag{11}$$

This follows from the constraint of any SRV to be ignorant about any individual variable in $X$, so $\sigma(X_1) = \varnothing$.

### 2.3.5 Zero synergy in a single variable

$$I_{\text{syn}}(X \to X_1) = 0. \tag{12}$$

This also follows from the constraint of any SRV to be ignorant about any individual variable in $X$: all terms in the sum in Eq. (6) are necessarily zero.

### 2.3.6 Identity maximizes synergistic information

$$I_{\text{syn}}(X \to X) = H(\Sigma(X)) = \max. \tag{13}$$

This follows from the fact that each $S_i \in \Sigma^\perp_{(S_i)_i}(X)$ is computed from $X$ and is therefore completely redundant given $X$, so each term in the sum in Eq. (6) must be maximal and equal to $H(S_i^\perp)$. Since all $S_i^\perp \in \Sigma^\perp_{(S_i)_i}(X)$ are independent, $\sum_i H(S_i^\perp) = H\left(\Sigma^\perp_{(S_i)_i}(X)\right)$.

## 2.4 Consequential properties

We now list important properties which are induced by our proposed synergy measure $I_{\text{syn}}(X \to Y)$ along with their corresponding proofs.

### 2.4.1 Upper bound on the mutual information of an SRV

The maximum amount of mutual information (and entropy) of an SRV of a set of variables can be derived analytically. We start with the case of two input variables, i.e., $|X| = 2$, and then generalize. Maximizing $I(X_1, X_2 : S)$ under the two constraints $I(X_1 : S) = 0$ and $I(X_2 : S) = 0$ from Eq. (3) leads to

$$\begin{aligned}
I(X_1, X_2 : S) &= I(X_1 : S) + I(X_2 : S \mid X_1), \\
&= I(X_2 : S \mid X_1), \\
&= H(X_2 \mid X_1) - H(X_2 \mid S, X_1), \\
&\leq H(X_2 \mid X_1).
\end{aligned} \tag{14}$$

using that $I(X_1 : S) = 0$ by constraint. Since the first term in the third line $H(X_2 \mid X_1)$ does not change by varying $S$ we can maximize $I(X_1, X_2 : S)$ only by minimizing the second term $H(X_2 \mid S, X_1)$. Since $H(X_2 \mid S, X_1) \geq 0$ and the same holds if $X_1$ and $X_2$ are swapped (reordering the labelling) we find that

$$I(S : X_1, X_2) \leq \min(H(X_2 \mid X_1), H(X_1 \mid X_2)), \text{ where } S \in \sigma(X_1, X_2). \tag{15}$$

This can be rewritten as

$$I(S : X_1, X_2) \leq H(X_1, X_2) - \max(H(X_1), H(X_2)), \text{ where } S \in \sigma(X_1, X_2). \tag{16}$$

The generalization to $N$ variables is fairly straightforward by induction (see Appendix 6.1) and here illustrated for the case $N = 3$ for one particular labeling $(X_i)_i$:

$$\begin{aligned}
I(X_1, X_2, X_3 : S) &= I(X_1, X_2 : S) + I(X_3 : S | X_1, X_2) \\
&= I(X_2 : S_1 | X_1) + I(X_3 : S | X_1, X_2) \\
&= [H(X_2 | X_1) - H(X_2 | S, X_1)] + [H(X_3 | X_1, X_2) - H(X_3 | S, X_1, X_2)] \\
&\leq H(X_2 | X_1) + H(X_3 | X_1, X_2) \\
&= H(X_1, X_2, X_3) - H(X_1).
\end{aligned}$$

(17)

Since this inequality must be true for all labelings of the $(X_i)_i$, in particular for the labeling that maximizes $H(X_1)$, and extending this result to any $N$, we find that

$$I(X_1, ..., X_N : S) \leq H(X_1, ..., X_N) - \max_i H(X_i), \text{ where } S \in \sigma(X). \qquad (18)$$

*Corollary*. Suppose that $Y$ is completely synergistic about $X$, i.e., $Y \in \sigma(X)$. Then their mutual information is bounded as follows:

$$I(X : Y) \leq H(X_1, ..., X_N) - \max_i H(X_i), \text{ where } Y \in \sigma(X). \qquad (19)$$

Finally, we assume that the SRV is 'efficient' in the sense that it contains no additional entropy that is unrelated to $X$, i.e., $I(X_1, X_2 : S) = H(S)$. After all, if it would contain additional entropy then by our orthogonal decomposition assumption we can distill only the dependent part exactly. Therefore the derived upper bound of any SRV is also the upper bound on its entropy.

2.4.2  Non-equivalence of SRVs

It is indeed possible to have at least two non-redundant MSRVs in $\Sigma(X)$, i.e., $I(S_1 : S_2) < H(S_1)$ where $S_1, S_2 \in \Sigma(X)$, or even $I(S_1 : S_2) = 0$. In words, this means that there can be multiple types of synergistic relation with $X$ which are not equivalent. This is demonstrated by the following example: $X = \{X_1, X_2\}$ with $X_i \in \{0, 1, 2\}$ and uniform distribution $\Pr(X) = 1/9$, where $S_1 \equiv (2 - X_1 + X_2) \bmod 3$ and $S_2 = (X_1 + X_2) \bmod 3$. The fact that these functions are MSRVs is verified numerically by trying all combinations. It can also be seen visually in Figure 1; adding additional states for $S_1$ or $S_2$ or changing their distribution will break the symmetries needed to stay uncorrelated with the individual inputs. In this case $I(S_1 : S_2) = 0$ so the two MSRVs are mutually independent, whereas $I(S_1 : X) = I(S_2 : X) = \log_2 3 \approx 1.58$. In fact, as shown in Section 2.4.1 this is actually the maximum possible mutual information that any SRV can store about $X$. Since the MSRVs are a subset of the SRVs it follows trivially that SRVs can be non-equivalent or even independent.

|     | $X_2$ |   |   |
|-----|---|---|---|
| $X_1$ | 2 | 0 | 1 |
|     | 1 | 2 | 0 |
|     | 0 | 1 | 2 |

$S_1(X_1, X_2)$

|     | $X_2$ |   |   |
|-----|---|---|---|
|     | 0 | 1 | 2 |
|     | 1 | 2 | 0 |
|     | 2 | 0 | 1 |

$S_2(X_1, X_2)$

*Figure 1: the values of the two MSRVs S1 and S2 which are mutually independent but highly synergistic about two 3-valued variables X1 and X2. X1 and X2 are uniformly distributed and independent.*

2.4.3  Synergy among MSRVs

The combination of two (or more) MSRVs $(S_1, S_2)$, $S_1, S_2 \in \Sigma(X)$, cannot be an SRV, i.e., $(S_1, S_2) \notin \sigma(X)$. Otherwise it would be a contradiction: if $(S_1, S_2) \in \sigma(X)$ would be true then it follows that $S_1, S_2 \notin \Sigma(X)$, since both $S_1$ and $S_2$ would be completely redundant given $(S_1, S_2)$, and therefore discarded in the construction of MSRV.

This means that all combinations of MSRVs, such as $(S_1, S_2)$, must necessarily have non-zero mutual information about at least one of the individual source variables, i.e., $\exists i : I(S_1, S_2 : X_i) > 0$, violating Eq. (3). Since each individual MSRV has zero mutual information with each individual source variable by definition, it must be true that this 'non-synergistic' information results from synergy among MSRVs. We emphasize that this type of synergy among the $S_i \in \Sigma(X)$ is different from the synergy among the $X_i \in X$ which we intend to quantify in this paper, and could more appropriately be considered as a 'synergy of synergies'.

The fact that multiple MSRVs are possible is already proven by the example used in the previous proof in Section 2.4.2. The synergy among these two MSRVs in this example is indeed easily verified: $I(S_1 : X_1) = 0$ and $I(S_2 : X_1) = 0$, whereas $I(S_1, S_2 : X_1) = I(S_1, S_2 : X_1) = \log_2 3$.

Since MSRVs are a subset of the SRVs it follows that also SRVs can have such 'synergy-of-synergies'. In fact, the existence of multiple MSRVs means that there are necessarily SRVs which are synergistic about another SRV, and conversely, if there is only one MSRV then there cannot be any set of SRVs which are synergistic about another SRV.

*Corollary.* Alternatively quantifying synergistic information using directly the mutual information $I\left(Y : \Sigma^{\perp}_{(S_i)_i}(X)\right)$ could violate the fourth desired property, "Zero synergy about a single variable", because if $\Sigma(X)$ consists of two or more MSRVs then $\exists i : I(S_1, S_2 : X_i) > 0$. In this case the choice $Y \equiv X_i$ would have non-zero synergistic information about $X$, which is undesired.

### 2.4.4 Each XOR-gate of random binary inputs is an MSRV

Lastly we use our definition of synergy to prove the common intuition that the XOR-gate is maximally synergistic about a set of i.i.d. binary variables (bits), as suggested in the introductory example.

We start with the case of two bits $X_1, X_2 \in \{0,1\}$. As SRV we take $S_1 \equiv X_1 \oplus X_2$. The entropy of this SRV equals 1, which is in fact the upper bound of any SRV for this $X$, Eq. (18). Therefore no other SRV can make $S_1$ completely redundant such that it would prevent $S_1$ from becoming an MSRV (Section 2.2.2). It is only possible for another SRV to make $S_1$ redundant in case the converse is also true, in which case the two SRVs are equivalent. An example of this would be the NXOR (NOT-XOR) gate which is informationally equivalent to XOR. Here we consider equivalent SRVs as one and the same.

For the more general case of $N$ bits $X_1, ..., X_N \in \{0,1\}$, consider as SRV the set of XOR-gates $S \equiv \{S_1, S_2, ..., S_{N-1}\}$ where $S_i \equiv X_1 \oplus ... \oplus X_{i+1}$. It is easily verified that $S$ does not contain mutual information about any individual bit $X_i$, so indeed $S \in \sigma(X)$. Moreover it is also easily verified that all $S_i$ are independent, so the entropy $H(S) = N - 1$ which equals the upper bound on any SRV. Following the same reasoning as the two-bit case, $S$ is indeed an MSRV. We remark that conversely, each possible set of XOR gates is not necessarily an MSRV because, e.g., $X_1 \oplus X_3$ is redundant given both $X_1 \oplus X_2$ and $X_2 \oplus X_3$. That is, some (sets of) XOR-gates are redundant given others and will therefore not be member of the set $\Sigma(X)$ by construction.

### 2.5 Numerical implementation

We have implemented the numerical procedures to compute the above as part of a Python library named *jointpdf* (available online[1]). Here, a set of discrete stochastic variables $X$ is represented by a matrix of joint probabilities of dimensions $m^n$, where $n$ is the number of variables and $m$ is the number of possible values per variable. This matrix is uniquely identified by $m^n - 1$ independent parameters each on the unit line.

In brief, finding an MSRV $S$ amounts to numerically optimizing a subset of the (bounded) parameters of $\Pr(X, S)$ in order to maximize $I(S:X)$ while satisfying the conditions for SRVs in Eq. (3). Then we approximate the set of OSRVs $\Sigma^\perp_{(S_i)_i}(X)$ by constructing it iteratively. For finding the next OSRV $S_N$ in addition to an existing set $S_1, ..., S_{N-1}$, the independence constraint $I(S_N : S_1, ..., S_{N-1}) = 0$ is added to the numerical optimization. The procedure finishes once no more OSRVs are found. The optimization of their ordering is implemented by restarting the sequence of numerical optimizations from different starting points and taking the result with highest synergistic information. Orthogonal decomposition is also implemented even though it is not used since the OSRV set is built directly using this optimization procedure. This uses the fact that each decomposed part of an SRV must also be an SRV (assuming perfect orthogonal decomposition) and can therefore be found directly in the optimization. For all numerical optimizations the algorithm *scipy.optimize.minimize* (version 0.11.0) is used.

Once the probability distribution is extended with the set of OSRVs, the amount of synergistic information has a confidence interval due to the approximate nature of the numerical optimizations.

---
[1] https://bitbucket.org/rquax/jointpdf

That is, one or more OSRVs may turn out to store a small amount of unwanted information about individual inputs. We subtract these unwanted quantities from each mutual information term in Eq. (6) in order to estimate the synergistic information in each OSRV. However, these subtracted terms could be (partially) redundant, the extent of which cannot be determined in general. Thus, once the optimal sequence of OSRVs is found we take the lower bound on the estimated synergistic information $\hat{I}_{\text{syn}}(X \to Y)$ as:

$$\hat{I}_{\text{syn}}(X \to Y) \geq \sum_{i \in 1,\ldots,N} \left[ I(S_i : X) - \sum_j I(S_i : X_j) \right].$$

This corresponds to the case where each subtracted mutual information term is fully independent so that they can be summed, leading to this WMS form [6]. On the other hand, the corresponding upper bound would occur if all subtracted mutual information terms would be fully redundant, in which case:

$$\hat{I}_{\text{syn}}(X \to Y) \leq \sum_{i \in 1,\ldots,N} \left[ I(S_i : X) - \max_j I(S_i : X_j) \right].$$

We take the middle point between these bounds as the best estimate $\hat{I}_{\text{syn}}(X \to Y)$. The corresponding measure of uncertainty is then defined as the *relative error* as

$$\Delta \hat{I}_{\text{syn}}(X \to Y) \equiv \frac{\sum_j I(S_i : X_j) - \max_j I(S_i : X_j)}{\hat{I}_{\text{syn}}(X \to Y)}. \tag{20}$$

The following numerical results have been obtained for the case of two input variables, $X_1$ and $X_2$, and one output variable $Y$. Their joint probability distribution $\Pr(X_1, X_2, Y)$ is randomly generated unless otherwise stated. Once an OSRV is found it is added to this distribution as an additional variable. All variables are constrained to have the same number of possible values ('state space') in our experiments.

### 2.5.1 Success rate and accuracy of finding SRVs

Our first result is on the ability of our numerical algorithm to find a single SRV as function of the number of possible states per individual variable. Namely, our definition of synergistic information in Eq. (6) relies on perfect orthogonal decomposition; we showed that perfect orthogonal decomposition is impossible for at least one type of relation among binary variables (Appendix 6.5), whereas previous work hints that continuous variables might be (almost) perfectly decomposed (Section 2.1.4). Figure 2 shows the probability of successfully finding an SRV for variables with a state space of 2, 3, 4 and 5 values. Success is defined as a relative error on the entropy of the SRV of less than 10%.

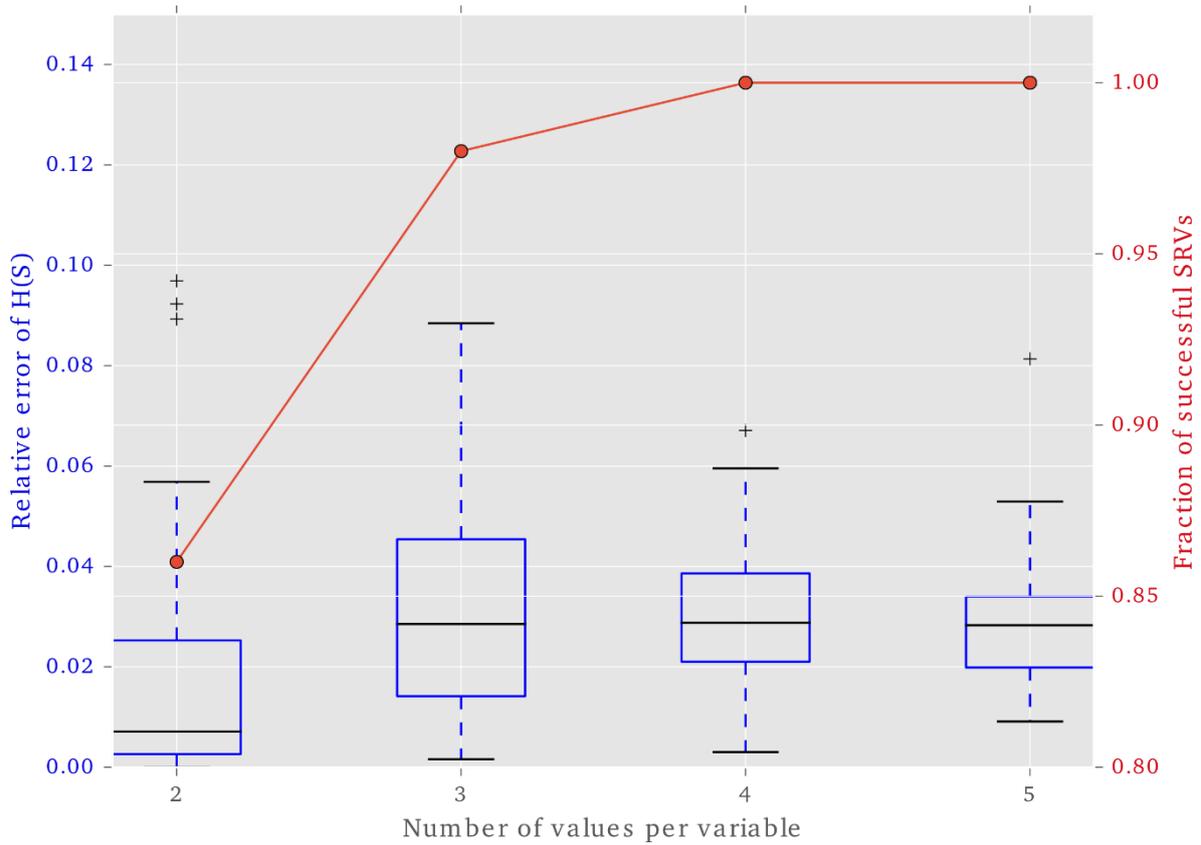

*Figure 2: Effectiveness of the numerical implementation to find a single SRV. The input consists of two variables with 2, 3, 4, or 5 possible values each (x-axis). Red line with dots: probability that an SRV could be found with at most 10% relative error in 50 randomly generated $\Pr(X_1, X_2, Y)$ distributions. The fact that it is lowest for binary variables is consistent with the observation that perfect orthogonal decomposition is impossible in this case under at least one known condition (Appendix 6.5). The fact that it converges to 1 is consistent with our suggestion that orthogonal decomposition could be possible for continuous variables (Section 2.1.4). Blue box plot: expected relative error of the entropy of a single SRV, once successfully found.*

In Figure 2 we also show the expected relative error on the entropy of an SRV once successfully found. This is relevant for our confidence in the subsequent results. For 2 or 3 values per variable we find a relative error in the low range of 1-3%, indicating that finding an SRV is a bimodal problem: either it is successfully found with relatively low error or it is not found successfully and has high error. For 4 or more values per variable a satisfactory SRV is always successfully found. This indicates that additional degrees of freedom aid in finding SRVs.

### 2.5.2 Efficiency of a single SRV

Once an SRV is successfully found, the next question is how much synergistic information it actually contains compared to the maximum possible. According to Eg. (18), the upper bound is the minimum of $H(X_2 | X_1)$ and $H(X_1 | X_2)$. Thus, a single added variable as SRV has in principle sufficient entropy to store this information. However, depending on $\Pr(X_1, X_2)$ it is possible that a single SRV cannot store all synergistic information at once, regardless of how much entropy it has, as demonstrated in Section 2.4.3. This happens if two or more SRVs would be mutually 'incompatible' (cannot be combined into a single, large SRV). Therefore we show the expected synergistic information in a single SRV normalized by the corresponding upper bound in Figure 3.

The decreasing trend indicates that this incompatibility among SRVs plays a significant role as the state space of the variables grows. This would imply that an increasing number of SRVs must be found in order to estimate the total synergistic information $\hat{I}_{\text{syn}}(X \rightarrow Y)$. Fortunately, Figure 3 also suggests that the efficiency settles to a non-zero constant which implies that the number of needed SRVs does not grow to impractical numbers.

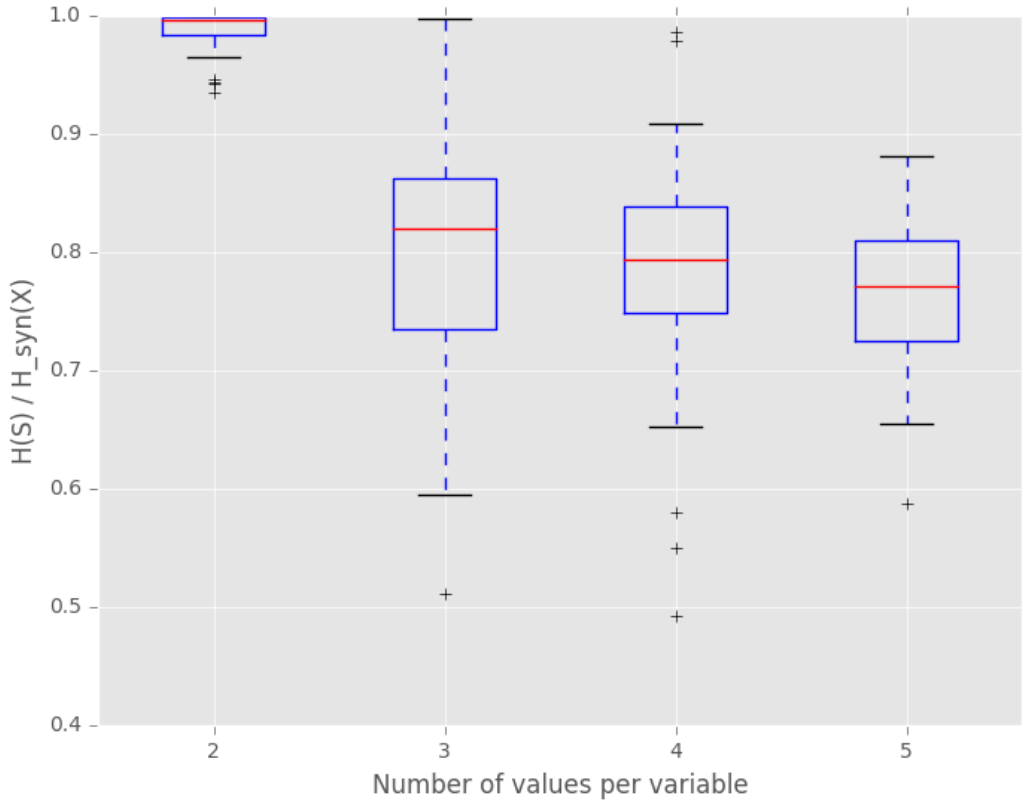

*Figure 3: Synergistic entropy of a single SRV normalized by the theoretical upper bound. The input consists of two randomly generated stochastic variables with 2, 3, 4, or 5 possible values per variable (x-axis). The SRV is constrained to have the same number of possible values. The initial downward trend shows that individual SRVs become less efficient in storing synergistic information as the state space per variable grows. The apparent settling to a non-zero constant suggests that estimating synergistic information does not require a diverging number of SRVs to be found for any number of values per variable.*

### 2.5.3 Resilience implication of synergy

Finally we compare the impact of two types of perturbations in two types of input-output relations, namely the case of a randomly generated $\Pr(Y|X_1, X_2)$ versus the case that $Y$ is an SRV of $X$. A 'local' perturbation is implemented by adding a random vector with norm 0.1 to the point in the unit hypercube that defines the marginal distribution of a randomly selected input variable, so $P(X_1)$ or $P(X_2)$. Conversely, a 'non-local' perturbation is similarly applied to $P(X_2|X_1)$ while keeping $P(X_1)$ and $P(X_2)$ unchanged. The impact is quantified by the relative change of the mutual information $I(X_1, X_2 : Y)$ due to the perturbation. That is, we ask whether a small perturbation disrupts the information transmission when viewing $X_1, X_2 \to Y$ as a communication channel.

In Figure 4 we show that a synergistic $Y$ is significantly less susceptible to local perturbations compared to a randomly generated $Y$. For non-local perturbations the difference in susceptibility is smaller but still significant. The null-hypothesis of equal population median is rejected both for local and non-local perturbations (Mood's median test, p-values $1.2 \cdot 10^{-13}$ and $5.5 \cdot 10^{-5}$ respectively; threshold $0.01$).

The difference in susceptibility for local perturbations is intuitive because an SRV has zero mutual information with individual inputs, so it is arguably insensitive to changes in individual inputs. We still find a non-zero expected impact; this could be partly explained by our algorithm's relative error being on the order of 3% which is the same order as the relative impact found (2%). In order to test this intuition we devised the non-local perturbations to compare against. A larger susceptibility is indeed found for non-local perturbations, however it remains unclear why synergistic variables are still less susceptible in the non-local case compared to randomly generated variables. Nevertheless, our numerical results indicate that synergy plays a significant role in resilience to noise. This is relevant especially for biological systems which are continually subject to noise and must be resilient to it.

A simple use-case on using the *jointpdf* package to estimate synergies, as is done here, is included in Appendix 6.7.

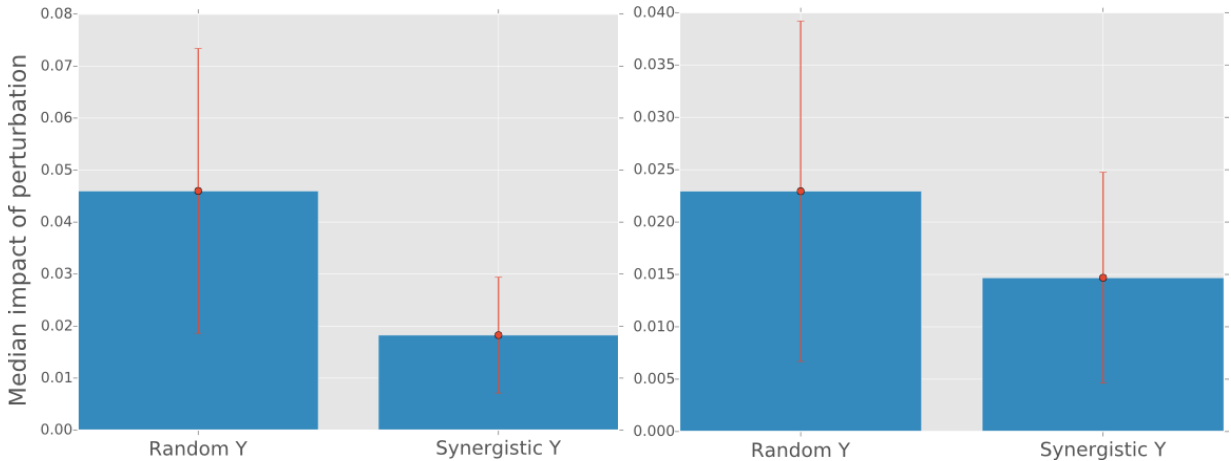

*Figure 4: Left: The median relative change of the mutual information $I(X_1, X_2 : Y)$ after perturbing a single input variable's marginal distribution $P(X_1)$ ('local' perturbation). Error bars indicate the 25th and 75th percentiles. A perturbation is implemented by adding a random vector with norm 0.1 to the point in unit hypercube that defines the marginal distribution $P(X_1)$. Each bar is based on 100 randomly generated joint distributions $P(X_1, X_2, Y)$, where in the synergistic case $Y$ is constrained to be an SRV of $X_1, X_2$. Right: the same as left except that the perturbation is 'non-local' in the sense that it is applied to $P(X_2|X_1)$ while keeping $P(X_1)$ and $P(X_2)$ unchanged.*

# 3 Discussion

Most theoretical work on defining synergistic information uses the PID framework [3], which (informally stated) requires that $I(X:Y)=\text{synergy}+\text{individual}$. That is, the more synergistic information $Y$ stores about $X$, the less information it can store about an individual $X_i$ because those two types of information are required to sum up to the quantity $I(X:Y)$ as non-negative terms. Our approach is incompatible with this viewpoint. That is, in our framework the amount of synergistic information $I_{\text{syn}}(X \to Y)$ makes no statement on the amount of 'individual' information that $Y$ may also store about $X_i$. In fact, the proposed synergistic information $I_{\text{syn}}(X \to Y)$ can be maximized by the identity $I_{\text{syn}}(X \to X)$, which obviously also stores maximum information about all individual variables $X_i$. The fact that no successful synergy measure has been found to date that satisfies the PID framework led us to explore a completely different viewpoint. If our proposed measure would prove successful then it may imply that the decomposition requirement is too strong for a synergy measure to obey. Whether our proposed synergy measure can be used to define a different notion of decomposition remains an open question.

We find an additional argument against the decomposition requirement in the example in Section 2.4.2 and 2.4.3. This example demonstrates that two independent SRVs (zero mutual information) can exist which are not synergistic when taken together. That is, there are evidently two distinct ways in which a variable $Y$ can be completely synergistic about $X$. However, it is impossible to store information about both these SRVs (maximum synergy) while still having zero information about all individual variables $X_i$. This suggests that synergistic information and 'individual' information cannot simply be considered as mutually exclusive.

Therefore we propose an alternative viewpoint. Whereas synergistic information could be measured by $I_{\text{syn}}(X \to Y)$, the amount of 'individual information' could foreseeably be measured by a similar procedure. For instance, the set $\Sigma^{\perp}_{(S_i)_i}$ could be replaced by the individual inputs $(X_i)_i$ after which the same procedure in Eq. (6) as for $I_{\text{syn}}(X \to Y)$ is repeated. This would measure the amount of 'unique' information that $Y$ stores about individual inputs which is not also stored in (combinations of) other inputs. This measure would be upper bounded by $H(X)$. For $N$ completely random and independent inputs, this individual information would be upper bounded by $N \cdot H(X_1)$ whereas the synergistic information would be upper bounded by $(N-1) \cdot H(X_1)$ (in any SRV). Both quantities measure different but not fully independent aspects. How the two measures relate to each other is subject of future work.

Our proposed definition builds upon the concept of orthogonal decomposition. It allows us to rigorously define a single, definite measure of synergistic information from first principles. However further research is needed to determine for which cases this decomposition can be done exactly, approximately, or not at all. Even if in a specific case it would turn out to be not exactly computable (due to imperfect orthogonal decomposition) then our definition can still serve as a reference point. To the extent that a necessary orthogonal decomposition must be approximated (or bounded), the resulting amount of synergistic information must also be considered an approximation (or bound).

Our final point of discussion is that the choice of how to divide a stochastic variable $X$ into subvariables $X \equiv \{X_i\}_i$ is crucial and determines the amount of information synergy found. This choice strongly depends on the specific research question. For instance, the neurons of a brain may be divided into the two cerebral hemispheres, into many anatomical regions, or into individual neurons altogether, where at each level the amount of information synergy may differ. In this article we are not concerned with choosing the division and will calculate the amount of information synergy once the subvariables have been chosen.

# 4 Conclusion

In this paper we propose a measure to uniquely quantify synergistic information from first principles. Briefly, we first 'extract' all synergistic entropy of a set of variables $X \equiv \{X_i\}_i$ by constructing a new set of all possible *maximally synergistic random variables* (MSRVs) of $X$, denoted $\Sigma(X)$, where each MSRV has non-zero mutual information with the set $X$ but zero mutual information with any individual $X_i$. This set of MSRVs is then transformed into a set of independent *orthogonal* SRVs (OSRV), denoted $\Sigma^{\perp}_{(S_i)_i}(X)$, to prevent over counting. Then we define the amount of synergistic information in outcome variable $Y$ about the set of source variables $X$ as the sum of OSRV-specific mutual information quantities, $\sum_{S_i \in \sigma^{\perp}_{(S_i)_i}(X)} I(S_i : Y)$.

Our proposed measure satisfies important desired properties, e.g., it is non-negative and bounded by mutual information, invariant to rearranging $X$, and always has zero synergy if the input is a single variable. We also prove four important properties of our synergy measure. In particular, we derive the maximum mutual information in case $Y$ is an SRV; we demonstrate that synergistic information can be of different types (multiple, independent SRVs); and we prove the fact that the combination of multiple SRVs may store non-zero information about an individual $X_i$ in a synergistic way. This latter property leads to the intriguing concept of 'synergy among synergies', which we show must necessarily be excluded from quantifying synergy in $Y$ about $X$ but which might turn out to be an interesting subject of study in its own right. Finally, we provide a software implementation of the proposed synergy measure.

The ability to quantify synergistic information in an arbitrary multivariate setting is a necessary step to better understand how dynamical systems implement their complex information processing capabilities. Our proposed framework based on SRVs and orthogonal decomposition provides a new line of thinking and produces a general synergy measure with important desired properties. Our initial numerical experiments suggest that synergistic relations are less sensitive to noise, which is an important property of biological and social systems. Studying the information synergy in complex adaptive systems will certainly lead to substantial new insights into their various emergent behaviors, ranging from

# 5 Acknowledgements

The authors acknowledge the financial support of the Future and Emerging Technologies (FET) program within Seventh Framework Programme (FP7) for Research of the European Commission, under the FET- Proactive grant agreement TOPDRIM, number FP7-ICT-318121, as well as the financial support of the Future and Emerging Technologies (FET) program within Seventh Framework Programme (FP7) for Research of the European Commission, under the FET-Proactive grant agreement

Sophocles, number FP7-ICT-317534. PMAS acknowledges the support of the Russian Scientific Foundation, Project number 14-21-00137.

# 6 Appendix

## 6.1 Upper bound of possible entropy of an SRV by induction

### 6.1.1 Base case

The base case is that it is true that $I(X_1,...,X_{N-1}:S) \leq H(X_1,...,X_{N-1}) - \max_{1 \leq i \leq N-1} H(X_i)$, which is proven for $N=2$ in Section 2.4.1.

### 6.1.2 Induction step

We will prove that the base case induces $I(X_1,...,X_N:S) \leq H(X_1,...,X_N) - \max_{1 \leq i \leq N} H(X_i)$.

An $S \in \sigma(X)$ must be chosen which maximizes:

$$\begin{aligned}
I(X_1,...,X_{N-1},X_N:S) &= I(X_1,...,X_{N-1}:S) + I(X_N:S|X_1,...,X_{N-1}) \\
&= H(X_1,...,X_{N-1}) - \max_{1 \leq i \leq N-1} H(X_i) + I(X_N:S|X_1,...,X_{N-1}) \\
&= H(X_1,...,X_{N-1}) - \max_{1 \leq i \leq N-1} H(X_i) + H(X_N|X_1,...,X_{N-1}) \\
&\quad - H(X_N|S,X_1,...,X_{N-1}) \\
&= H(X_1,...,X_N) - \max_{1 \leq i \leq N-1} H(X_i) - H(X_N|S,X_1,...,X_{N-1}) \\
&\leq H(X_1,...,X_N) - \max_{1 \leq i \leq N-1} H(X_i).
\end{aligned}$$

Here, the maximization term is understood to maximize over all label orderings of the first $N-1$ subvariables $\{X_1,...,X_{N-1}\}$. We note that this upper bound relation must be true for all choices of orderings $\{X_i\}_{1 \leq i \leq N}$ of all $N$ labels (since the labeling is arbitrary and due to the desired property in Section 2.3.3). Therefore, $S$ must satisfy all $N!$ simultaneous instances of the above inequality, one for each possible ordering. Any $S$ that satisfies the 'most constraining' inequality, i.e., where the r.h.s. is minimal, necessarily also satisfies all $N!$ inequalities. The r.h.s. is minimized in case the $X_i$ with overall maximum $H(X_i)$ is part of the subset $\{X_1,...,X_{N-1}\}$. In other words, for the inequality with minimal r.h.s. it is true that

$$\max_{1 \leq i \leq N-1} H(X_i) = \max_{1 \leq i \leq N} H(X_i).$$

Substituting this above we find indeed that

$$I(X_1,...,X_N:S) \leq H(X_1,...,X_N) - \max_{1 \leq i \leq N} H(X_i).$$

## 6.2 $I_{\text{syn}}(X \to Y)$ does not 'overcount' any synergistic information

All synergistic information that any $Y$ can store about $X$ is encoded by the set of SRVs $\sigma(X)$, which is informationally equivalent to $\sigma^{\perp}_{(S_i)_i}(X)$, i.e., they have equal entropy and zero conditional entropy. Therefore $I(Y:\sigma^{\perp}_{(S_i)_i}(X))$ should be an upper bound on $I_{\text{syn}}(X \to Y)$ since otherwise some synergistic information must have been doubly counted. In this section we derive that

$I_{\text{syn}}(X \to Y) \leq I\left(Y : \sigma_{(S_i)_i}^{\perp}(X)\right)$. In Appendix 6.2.1 we use the same derivation to demonstrate that a positive difference $I\left(Y : \sigma_{(S_i)_i}^{\perp}(X)\right) - I_{\text{syn}}(X \to Y)$ is undesirable at least in some cases.

Here we start with the proof that $I_{\text{syn}}(X \to Y) \leq I\left(Y : \sigma_{(S_i)_i}^{\perp}(X)\right)$ in case $\sigma_{(S_i)_i}^{\perp}(X)$ consists of two OSRVs, taken as base case $n=2$ for a proof by induction. Then we also work out the case $n=3$ so that the reader can see how the derivation extends for increasing $n$. Then we provide the proof by induction in $n$.

Let $\sigma_{(S_i)_i}^{\perp}(X) = S_1, \ldots, S_N$ consist of an arbitrary number $N$ of OSRVs. Let $\sigma_{(S_i)_i}^{\perp}(X)_n \equiv S_1, \ldots, S_n$ denote the first $n$ OSRVs for $n \leq N$. Let $I_{\text{syn}}(X \to Y)_n$ be defined using $\sigma_{(S_i)_i}^{\perp}(X)_n$ instead of $\sigma_{(S_i)_i}^{\perp}(X)$, i.e., only the first $n$ terms in the sum in Eq. (6).

For $n=2$ we use the property $H(S_1|S_2) = H(S_1)$ by construction of $\sigma_{(S_i)_i}^{\perp}(X)$:

$$
\begin{aligned}
I\left(Y : \sigma_{(S_i)_i}^{\perp}(X)_{n=2}\right) &= I(Y : S_1, S_2) \\
&= I(Y : S_1) + I(Y : S_2 | S_1) \\
&= I(Y : S_1) + I(Y : S_2) + H(S_1|Y) + H(S_1|S_2) - H(S_1|Y, S_2) - H(S_1) \\
&= I(Y : S_1) + I(Y : S_2) + H(S_1|Y) - H(S_1|Y, S_2) \\
&\geq I(Y : S_1) + I(Y : S_2) \\
&= I_{\text{syn}}(X \to Y)_{n=2}.
\end{aligned}
$$

For $n=3$ we similarly use the independence properties $H(S_1|S_2) = H(S_1)$ and $H(S_1, S_2 | S_3) = H(S_1, S_2)$:

$$
\begin{aligned}
I\left(Y : \sigma_{(S_i)_i}^{\perp}(X)_{n=3}\right) &= I(Y : S_1, S_2, S_3) \\
&= I(Y : S_1) + I(Y : S_2 | S_1) + I(Y : S_3 | S_1, S_2) \\
&= I(Y : S_1) + I(Y : S_2) + H(S_1|Y) + H(S_1|S_2) - H(S_1|Y, S_2) - H(S_1) \\
&\quad + I(Y : S_3 | S_1, S_2) \\
&= I(Y : S_1) + I(Y : S_2) + H(S_1|Y) - H(S_1|Y, S_2) + I(Y : S_3 | S_1, S_2) \\
&= I(Y : S_1) + I(Y : S_2) + I(Y : S_3) + H(S_1|Y) - H(S_1|Y, S_2) \\
&\quad + H(S_1, S_2|Y) - H(S_1, S_2|Y, S_3) - H(S_1, S_2) + H(S_1, S_2|S_3) \\
&= I_{\text{syn}}(X \to Y)_{n=3} + \left[H(S_1|Y) - H(S_1|Y, S_2)\right] + \left[H(S_1, S_2|Y) - H(S_1, S_2|Y, S_3)\right] \\
&\geq I_{\text{syn}}(X \to Y)_{n=3}.
\end{aligned}
$$

Essentially, the proof for each $n$ proceeds by rewriting each conditional mutual information term as a mutual information term plus four entropy terms (third equality above) of which two cancel out ($H(S_1, S_2) = H(S_1, S_2 | S_3)$ above) and the remaining two terms summed are non-negative ($H(S_1, S_2 | Y) \geq H(S_1, S_2 | Y, S_3)$ above). Thus, by induction:

$$\begin{aligned}
I\left(Y : \sigma^{\perp}_{(S_i)_i}(X)_n\right) &= I(Y : S_1, \ldots, S_n) \\
&= I(Y : S_1, \ldots, S_{n-1}) + I(Y : S_n | S_1, \ldots, S_{n-1}) \\
&= I(Y : S_1, \ldots, S_{n-1}) + I(Y : S_n) + H(S_1, \ldots, S_{n-1} | Y) - H(S_1, \ldots, S_{n-1} | Y, S_n) \\
&\quad - H(S_1, \ldots, S_{n-1}) + H(S_1, \ldots, S_{n-1} | S_n) \\
&= I(Y : S_1, \ldots, S_{n-1}) + I(Y : S_n) + \left[ H(S_1, S_2 | Y) - H(S_1, S_2 | Y, S_3) \right] \\
&\geq I_{\text{syn}}(X \to Y)_{n-1} + I(Y : S_n) + \left[ H(S_1, S_2 | Y) - H(S_1, S_2 | Y, S_3) \right] \\
&\geq I_{\text{syn}}(X \to Y)_n.
\end{aligned}$$

Thus we find that it is not possible for our proposed $I_{\text{syn}}(X \to Y)$ to exceed the mutual information $I\left(Y : \sigma^{\perp}_{(S_i)_i}(X)\right)$. This suggests that $I_{\text{syn}}(X \to Y)$ does not 'overcount' any synergistic information.

### 6.2.1 $I\left(Y : \sigma^{\perp}_{(S_i)_i}(X)\right)$ also includes non-synergistic information

In the derivation of the previous section we observe that, conversely, $I\left(Y : \sigma^{\perp}_{(S_i)_i}(X)\right)$ can exceed $I_{\text{syn}}(X \to Y)$ and we will now proceed to show that this is undesirable at least in some cases.

The positive difference $I\left(Y : \sigma^{\perp}_{(S_i)_i}(X)\right) - I_{\text{syn}}(X \to Y)$ must arise from one of the non-negative terms in square brackets in all derivations above. Suppose that $Y = X_i$ and therefore has zero information with any individual OSRV by definition (Eq. (section-2.2.1)). That is, $Y = X_i$ does not correlate with any possible synergistic relation (SRV) about $X$. In our view, $Y = X_i$ should thus be said to store zero synergistic information about $X$. However, even though $\forall i : H(S_i | Y) = H(S_i)$ by construction, this does not necessarily imply $H(S_1, \ldots, S_{n-1} | Y) = H(S_1, \ldots, S_{n-1} | Y, S_n)$, among others, and therefore any term in square brackets above can still be positive. In other words, it is possible for $Y = X_i$ to 'cooperate' or have synergy with one or more OSRVs to have non-zero mutual information about another OSRV. A concrete example of this is given in Section 2.4.3. This would lead to a non-zero synergistic information if quantified by $I\left(Y : \sigma^{\perp}_{(S_i)_i}(X)\right)$, which is undesirable in our view. In contrast, our proposed definition for $I_{\text{syn}}(X \to Y)$ in Eq. (6) purposely ignores this 'synergy-of-synergies' and in fact will always yield $I_{\text{syn}}(X \to Y) = 0$ in case $Y \equiv X_i$, which is desirable in our view and proved in Section 2.3.5.

## 6.3 Synergy measure correctly handles synergy-of-synergies among SRVs

By 'correctly handled' we mean that synergistic information is neither overcounted nor undercounted. We already start by the conjecture that 'non-synergistic' redundancy among a pair of SRVs does not lead to under or overcounting synergistic information. That is, suppose that $I(S_1 : S_2) > 0$, which we consider 'non-synergistic' mutual information. If $Y$ correlates with one or neither SRV then the optimal ordering is trivial. If it correlates with both then any ordering will do, assuming that their respective 'parallel' parts (see Section 2.2.5) are informationally equivalent and it does not matter which one is retained in $\sigma^{\perp}_{(S_i)_i}(X)$. The respective orthogonal parts are retained in any case. Therefore we now proceed to handle the case where there is synergy among SRVs.

First we illustrate the apparent problem which we handle in this section. Suppose that $\sigma(X) = \{S_1, S_2, S_3\}$ and further suppose that $I(S_1, S_2 : S_3) = H(S_3)$ while $\forall i, j : I(S_i : S_j) = 0$. In other words, by this construction the pair $S_1, S_2$ synergistically makes $S_3$ fully redundant, and no non-synergistic redundancy among the SRVs exists. Finally, let $S_3 \in Y$. At first sight it appears possible that $\sigma^{\perp}_{(S_i)_i}(X)$ happens to be constructed using an ordering $(S_i)_i$ such that $S_3$ appears after $S_1$ and $S_2$. This is unwanted because then $S_3$ will not be part of the $\sigma^{\perp}_{(S_i)_i}(X)$ used to compute $I_{\text{syn}}(X \to Y)$, i.e., the term $I(Y : S_3)$ disappears from the sum, which potentially leads to the constristution of $S_3$ to the synergistic information being ignored.

In this Appendix we show that the contribution is always counted towards $I_{\text{syn}}(X \to Y)$ by construction, and that the only possibility for the individual term $I(Y : S_3)$ to disappear is if its synergistic information is already accounted for.

First we interpret each such (synergistic) mutual information from a set of SRVs to another, single SRV as a ($n-1$ to 1) hyperedge in a hypergraph. In the above example, there would be a hyperedge from the pair $S_1, S_2$ to $S_3$. Let the weight of this hyperedge be equal to the mutual information. In the subsection 6.3.1 below we prove that in this setting, one hyperedge from $n-1$ SRVs to one SRV implies a hyperedge from all other possible $n-1$ subsets to the remaining SRV, at the same weight. That is, the hypergraph for $\sigma(X) = \{S_1, S_2, S_3\}$ forms a fully connected 'clique' of three hyperedges.

In this setting, finding a 'correct' ordering translates to letting $S_n$ appear before all $S_1, ..., S_{n-1}$ have appeared in case there is a hyperedge $S_1, ..., S_{n-1} \to S_n$ and $I(Y : S_n) > 0$. This translates to traversing a path of $n$ steps through the hyperedges in reverse order, each time choosing one SRV from the ancestor set that is not already previously chosen, such that for each SRV either (*i*) not all ancestor SRVs were chosen, or (*ii*) it has zero mutual information with $Y$. In other words, in case there is a $S_i$ such that $I(Y : S_i) = 0$ then any ordering with $S_i$ as last element will suffice. Only if $Y$ correlates with all SRVs then one of the SRVs will be (partially) discarded by the order maximization process in $I_{\text{syn}}(X \to Y)$. This is desirable because otherwise $I_{\text{syn}}(X \to Y)$ could exceed $I(X : Y)$ or even $H(Y)$. Intuitively, if $Y$ correlates with $n-1$ SRVs then it automatically correlates with the $n^{\text{th}}$ SRV as well, due to the redundancy among the SRVs. Counting this synergistic information would be overcounting this redundancy, leading to the violation of the boundedness by mutual information.

An example that demonstrates this phenomenon is given by $X \equiv \{X_1, X_2, X_3\}$ consisting of three i.i.d. binary variables. It has four pairwise-independent MSRVs, namely the three pairwise XOR functions and one nested 'XOR-of-XOR' function (verified numerically). However, one pairwise XOR is synergistically fully redundant given the two other pairwise XORs, so the entropy $H(\sigma(X)) = 3$, which equals $H(X)$. Taking e.g. $Y \equiv X$ yields indeed 3 bits of synergistic information according to our proposed definition of $I_{syn}(X \to Y)$, correctly discarding the synergistic redundancy among the four SRVs. However, if the synergistically redundant SRV would not be discarded from the sum then we would find 4 bits of synergistic information in $Y$ about $X$, which is counterintuitive because it exceeds $H(X)$, $H(Y)$, and $I(X:Y)$. Intuitively, the fact that $Y$ correlates with two pairwise XORs necessarily implies that it also correlates with the third pairwise XOR, so this redundant correlation should not be counted.

### 6.3.1 Synergy among SRVs forms a clique

Given is a particular set of SRVs $\sigma(X)$ in arbitrary order. Suppose that the set $S_1, S_2$ is fully synergistic about $S_3$, i.e., $I(S_1, S_2 : S_3) = d > 0$ and we first assume that $\forall i, j : I(S_i : S_j) = 0$. This assumption is dropped in the subsection below. The question is: are $S_2, S_3$ then also synergistic about $S_1$, and $S_1, S_3$ about $S_2$? We will now prove that in fact they are indeed synergistic at exactly the same amount, i.e., $I(S_2, S_3 : S_1) = I(S_1, S_3 : S_2) = d$. The following proof is thus for the case of two variables being synergistic about a third, but trivially generalizes to $n$ variables (in case the condition $\forall i, j : I(S_i : S_j) = 0$ is also generalized for $n-1$ variables).

First we find that the given condition $I(S_1, S_2 : S_3) = d > 0$ leads to known quantities for two conditional mutual information terms:

$$I(S_1, S_2 : S_3) = I(S_1 : S_3) + I(S_2 : S_3 | S_1) = I(S_2 : S_3) + I(S_1 : S_3 | S_2) = d$$
$$= I(S_2 : S_3 | S_1) = I(S_1 : S_3 | S_2) = d.$$

Then we use this to derive a different combination $I(S_1, S_3 : S_2)$ (the third combination is derived similarly):

$$I(S_1, S_3 : S_2) = I(S_1 : S_2) + I(S_3 : S_2 | S_1)$$
$$= I(S_3 : S_2) + H(S_1 | S_3) + H(S_1 | S_2) - H(S_1 | S_3, S_2) - H(S_1)$$
$$= -(H(S_1) - H(S_1 | S_3)) + (H(S_1 | S_2) - H(S_1 | S_2, S_3))$$
$$= -I(S_1 : S_3) + I(S_1 : S_3 | S_2)$$
$$= d.$$

In conclusion, we find that if a set of SRVs $S_1, \ldots, S_{n-1}$ synergistically stores mutual information about $S_n$ at amount $d$, then all subsets of $n-1$ SRVs of $S_1, \ldots, S_n$ will store exactly the same synergistic information about the respective remaining SRV. If each such synergistic mutual information from a set of SRVs to another SRV is considered as a directed ($n-1$ to 1) hyperedge in a hypergraph, then the resulting hypergraph of SRVs will have a clique in $S_1,...,S_n$.

### 6.3.2 Generalize to partial synergy among SRVs

Above we assumed $\forall i, j: I(S_i : S_j) = 0$. Now we remove this constraint and thus let all mutual informations of 2 (or $n-1$ in general) to be arbitrary. We then proceed as above, first:

$$I(S_1, S_2 : S_3) = I(S_1 : S_3) + I(S_2 : S_3 | S_1) = I(S_2 : S_3) + I(S_1 : S_3 | S_2) = d$$

Then:

$$\begin{aligned}
I(S_1, S_3 : S_2) &= I(S_1 : S_2) + I(S_3 : S_2 | S_1) \\
&= I(S_1 : S_2) + I(S_3 : S_2) + H(S_1 | S_3) + H(S_1 | S_2) - H(S_1 | S_3, S_2) - H(S_1) \\
&= I(S_1 : S_2) + I(S_3 : S_2) - (H(S_1) - H(S_1 | S_3)) + (H(S_1 | S_2) - H(S_1 | S_2, S_3)) \\
&= I(S_1 : S_2) + I(S_3 : S_2) - I(S_1 : S_3) + I(S_1 : S_3 | S_2) \\
&= I(S_1 : S_2, S_3) + I(S_3 : S_2) - I(S_1 : S_3) \\
&= d + I(S_3 : S_2) - I(S_1 : S_3).
\end{aligned}$$

We see that again $d$ is obtained for the mutual information among $n$ variables, but a correction term appears to account for a difference in the mutual information quantities among $n-1$ variables.

### 6.4 Independence of the two decomposed parts

From the first constraint $I(B^\perp : A) = 0$ it follows that:

$$\begin{aligned}
I(B^\perp : A) &= \sum_{a,b,b^\perp} p(a) p(b|a) p(b^\perp | b) \log \frac{p(a) p(b|a) p(b^\perp | b)}{p(a) p(b^\perp)} \\
&= \sum_{a,b,b^\perp} p(a) p(b|a) p(b^\perp | b) \log \frac{p(b|a) p(b^\perp | b)}{p(b^\perp)} \\
&= \sum_{a,b,b^\perp} p(a) p(b|a) p(b^\perp | b) \log \frac{p(b^\perp | b)}{p(b^\perp)} + \sum_{a,b,b^\perp} p(a) p(b|a) p(b^\perp | b) \log p(b|a) \\
&= \sum_{a,b,b^\perp} p(b) p(b^\perp | b) \log \frac{p(b^\perp | b)}{p(b^\perp)} + \sum_{a,b,b^\perp} p(a) p(b|a) \log p(b|a) \\
&= I(B^\perp : B) - H(B|A) = 0.
\end{aligned}$$

(21)

Here we used the shorthand $p(b|a) \equiv \Pr(B = b | A = a)$. From the resulting $I(B^\perp : B) = H(B|A)$ combined with the second constraint $I(B^\| : B) = I(B : A) = H(B) - H(B|A)$ it follows that $B^\perp$ and $B^\|$ must be independent, namely:

$$I(B^\perp, B^\| : B) = I(B^\perp : B) + I(B^\| : B | B^\perp)$$
$$= I(B^\perp : B) + I(B^\| : B) + H(B^\perp | B^\|) + H(B^\perp | B) - H(B^\perp | B^\|, B) - H(B^\perp)$$
$$= I(B^\perp : B) + I(B^\| : B) + H(B^\perp | B^\|) + H(B^\perp | B) - H(B^\perp)$$
$$= H(B)$$
$$= I(B^\perp : B) + I(B^\| : B).$$
(22)

6.5  Impossibility of decomposition for binary variables

Consider $A, B, B^\perp \in \{0,1\}$ as stochastic binary variables. The orthogonal decomposition $D : A, B \to B^\perp, B^\|$ imposes constraints on $B^\perp$ and $B^\|$ which cannot always be satisfied perfectly for the binary case, as we show next. We use the following model for $A$ and $B$:

$$p(A=1) \equiv p_a,$$
$$p(B=x|A=x) \equiv p_b.$$

In particular, we will show that $B^\perp$ cannot be computed from $B$ without storing information about $A$, violating the orthogonality condition. Being supposedly independent from $A$, we encode $B^\perp$ by its dependence on $B$ fully encoded by two parameters as

$$p(B^\perp = x | B = x) \equiv p_c^x. \tag{23}$$

Intuitively, in the case of binary variables, $B^\perp$ cannot store information about $B$ without also indirectly storing information about $A$. A possible explanation is that the binary case has an insufficient number of degrees of freedom for this.

To satisfy the condition $I(B^\perp : A) = 0$ it must be true that $\Pr(B^\perp | A) = \Pr(B^\perp)$ and therefore that $\Pr(B^\perp = 1 | A = 1) = \Pr(B^\perp = 1)$, among others. Let us find the conditions for this equality.

$$\Pr(B^\perp = 1 | A = 1) = \Pr(B^\perp = 1),$$
$$p_b p_c^1 + (1-p_b)(1-p_c^0) = p_a \left( p_b p_c^1 + (1-p_b)(1-p_c^0) \right) + (1-p_a)\left( (1-p_b) p_c^1 + p_b(1-p_c^0) \right),$$
$$1 = p_a + (1-p_a) \frac{(1-p_b) p_c^1 + p_b(1-p_c^0)}{p_b p_c^1 + (1-p_b)(1-p_c^0)},$$
$$p_b p_c^1 + (1-p_b)(1-p_c^0) = (1-p_b) p_c^1 + p_b(1-p_c^0).$$
(24)

The conditions for satisfying this equality are either $p_b = 1/2$ or $p_c^1 = 1 - p_c^0$. The first condition describes the trivial case where $B$ is independent from $A$. The second condition is less trivial but severely constrains the relations that $B^\perp$ can have with $A$. In fact it constrains the mutual information $I(B : B^\perp)$ to exactly zero regardless of $p_a$ and $p_b$, as we show next. Using the shorthand $p(B=1) = p_a p_b + (1-p_a)(1-p_b)$,

$$I\left(B^{\perp}:B\right) = \left(p(B=1)(p_c^0 + p_c^1 - 1) - p_c^0\right)\log\left(p_c^0 - p(B=1)(p_c^0 + p_c^1 - 1)\right)$$
$$-\left(p(B=1)(p_c^0 + p_c^1 - 1) - p_c^0 + 1\right)\log\left(p(B=1)(p_c^0 + p_c^1 - 1) - p_c^0 + 1\right)$$
$$+\left(p(B=1) - 1\right)\left((p_c^0 - 1)\log(1 - p_c^0) - p_c^0 \log(p_c^0)\right)$$
$$-p(B=1)\left((p_c^1 - 1)\log(1 - p_c^1) - p_c^1 \log(p_c^1)\right).$$
(25)

Using the substitution $p_c^1 = 1 - p_c^0$ and after some algebra steps it can be verified that indeed $I\left(B:B^{\perp}\right)$ simplifies to zero.

Extending the parameters $p_c^x$ to also depend on $A$ would certainly be possible and add degrees of freedom, however this can only create a non-zero conditional mutual information, $I\left(B:B^{\perp}|A\right)$. As soon as $I\left(B:B^{\perp}\right)$ is calculated then these extra parameters will be summed out into certain $p_c^x$ parameters, which we demonstrated will lead to zero mutual information under the orthogonality constraint.

This result demonstrates that a class of correlated binary variables $A$ and $B$ exists for which perfect orthogonal decomposition is impossible. Choices for binary $A$ and $B$ for which decomposition is indeed possible do exist, such as the trivial independent case. Exactly how numerous such cases are is currently unknown, especially when the number of possible states per variable is increased.

6.6 Wyner's common variable $B_W^{\parallel}$ satisfies orthogonal decomposition if $I\left(B_W^{\parallel}:A,B\right) = I\left(A:B\right)$.

Wyner's common variable is defined as a non-trivial minimization procedure, namely $B_W^{\parallel} = \arg\min_{A \to W \to B} I(A,B:W)$ where $A \to W \to B$ means that the minimization considers only random variables $W$ which make $A$ and $B$ independent, i.e., $I\left(A:B|W\right) = 0$. Wyner showed that in general $I\left(B_W^{\parallel}:A\right) \geq I\left(B:A\right)$ [13]. Here we show that for cases where the equality condition is actually reached, $B_W^{\parallel}$ satisfies all three orthogonal decomposition conditions which do not involve also $B^{\perp}$. Wyner leaves $B^{\perp}$ undefined and therefore his work cannot satisfy those conditions, but this shows at least one potential method of computing $B^{\parallel}$.

The two starting conditions are

$$I\left(B_W^{\parallel}:A,B\right) = I\left(A:B\right),$$
$$I\left(A:B|B_W^{\parallel}\right) = 0.$$
(26)

From the second condition it follows that

$$I\left(A:B,B_W^{\parallel}\right) = I\left(A:B_W^{\parallel}\right) + I\left(A:B|B_W^{\parallel}\right);$$
$$I\left(A:B,B_W^{\parallel}\right) = I\left(A:B_W^{\parallel}\right).$$
(27)

Similarly,

$$I(B:A,B_W^{\|}) = I(B:B_W^{\|}) + I(B:A|B_W^{\|});$$
$$I(B:A,B_W^{\|}) = I(B:B_W^{\|}).$$
(28)

Then from the first condition we can derive

$$\begin{aligned}
I(B_W^{\|}:A,B) &= I(A,B), \\
&= I(B:B_W^{\|}) + I(B_W^{\|}:A|B), \\
&= I(B:A,B_W^{\|}) + I(B_W^{\|}:A|B), \\
&= I(A:B) + I(B:B_W^{\|}|A) + I(B_W^{\|}:A|B),
\end{aligned}$$
(29)

from which follows

$$\begin{aligned}
I(B:B_W^{\|}|A) + I(B_W^{\|}:A|B) &= 0; \text{ so} \\
I(B:B_W^{\|}|A) &= 0, \\
I(B_W^{\|}:A|B) &= 0.
\end{aligned}$$
(30)

Firstly this implies the "non-spuriousness" condition on the last line. Then from combining Eq. (29) and Eq. (30) with either Eq. (27) or Eq. (28) we find, respectively,

$$\begin{aligned}
I(A:B) &= I(A:B_W^{\|}), \\
I(A:B) &= I(B:B_W^{\|}).
\end{aligned}$$
(31)

These are the "parallel" and "parsimony" conditions, concluding the proof.

## 6.7 Use-case of estimating synergy using the provided code

Our code can be run using any Python interface. As an example, suppose that a particular probability distribution is given of two 'input' stochastic variables, each having three possible values. We generate a random probability distribution as follows.

```
from jointpdf import JointProbabilityMatrix

# randomly generated joint probability mass function p(A,B)
# of 2 discrete stochastic variables, each having 3 possible values
p_AB = JointProbabilityMatrix(2,3)
```

We add a fully redundant (fully correlated) output variable as follows.

```
# append a third variable C which is deterministically computed from A and B, i.e., such
that I(A,B:C)=H(C)
p_AB.append_redundant_variables(1)
p_ABC = p_AB  # rename for clarity
```

Finally we compute the synergistic information $I_{\text{syn}}(A,B \to C)$ with the following command:

```
# compute the information synergy that C contains about A and B
p_ABC.synergistic_information([2], [0,1])
```

With the *jointpdf* package it is also easy to marginalize stochastic variables out of a joint distribution, add variables using various constraints, compute various information-theoretic quantities, and estimate distributions from data samples. It is implemented for discrete variables only. More details can be found on its website https://bitbucket.org/rquax/jointpdf.